\documentclass[12pt,preprint]{aastex}
\begin{document}

\title{On the Progenitor of Supernova 1987A}

\author{M. Parthasarathy\altaffilmark{1}, David Branch, E. Baron, and
  David J. Jeffery} \affil{Homer L. Dodge Department of Physics and
  Astronomy, University of Oklahoma, Norman, OK 73019}

\altaffiltext{1}{Indian Institute of Astrophysics, Koramangala,
  Bangalore - 560034, India}

\begin{abstract}

A previously unpublished ultralow--dispersion spectrum of Sanduleak
-69~202, the stellar progenitor of SN~1987A, is presented and the
uncertain presupernova evolution of Sanduleak -69~202 is discussed.

\end{abstract}


\section{INTRODUCTION}

Supernova 1987A in the Large Magellanic Cloud (LMC), the
observationally brightest supernova since Kepler's supernova of 1604,
was observed in all wavelength regions and its neutrino burst was
detected.  Its aftermath may be observed as long as there are
astronomers on Earth.  SN~1987A was classified as a Type~II supernova
(SN~II) in view of the strong hydrogen lines in its optical spectrum,
but because it was the explosion of a blue supergiant (BSG) rather
than a red one (RSG), it was an atypical SN~II: its light curve did
not reach maximum until three months after core collapse and at
maximum it was only about 10~percent as luminous as most SNe~II.  For
reviews of SN~1987A see Arnett et~al. (1989) and McCray (1993).
Vainu Bappu Observatory, Kavalur, India
 also participated in the observations of
SN~1987A (Ashoka et al. 1987, Anupama et al. 1988).

SN~1987A provided the first good opportunity to directly investigate a
supernova progenitor star.  The progenitor of SN~1987A was Sanduleak
-69~202 (Sk~-69~202), a twelfth magnitude star classified as OB by
Sanduleak (1970) from an objective prism survey of the LMC.
Unfortunately, the observational record of Sk~-69~202 is limited,
partly because no spectral peculiarities or variability called
attention to Sk~-69~202 during the decades prior to its demise.

Only serendipitous or survey data exist for Sk~-69~202.  Isserstedt
(1975) obtained UBV photoelectric photometry between 1971 and 1973: $V
= 12.24, B-V = +0.04$, and $U-B = -0.65$, characteristic of spectral
type B3~I.  Walborn et~al. (1987) analyzed eight CTIO 4--m
prime--focus plates obtained between 1974 and 1983 that covered the
wavelength range from blue to near--IR.  Sk~-69~202 appeared to be
normal, and no near--IR excess (no hot dust) was detected. Also, there
was little or no light variability between 1974 and 1983 (Walborn
et~al. 1987) and between 1970 and 1981 (Blanco et~al. 1987).  Plotkin
\& Clayton (2004) examined a large number of Harvard blue patrol
plates obtained from 1896 to 1954 and found no variability at the 0.3
magnitude level.

In 1972 and 1973 Rousseau et al. (1978) obtained ESO astrograph
objective--prism spectra of 110~\AA~mm$^{-1}$ dispersion at
H${\gamma}$ on Kodak IIa--O plates; the wavelength coverage was
limited, from 3850 to 4150~\AA. They classified Sk~-69~202 as B3~I.
Walborn et~al. (1989) analyzed and critically discussed these spectra
(as well as one ESO Schmidt objective prism spectrum of
460~\AA~mm$^{-1}$ dispersion at H${\gamma}$). They compared the
spectrum of Sk~-69~202 with the spectra of six other LMC supergiants
observed in the same way.  The spectral type of Sk~-69~202 was
constrained to be within the range B0.7 to B3, with B3 favored.

In view of the limited amount of observational data on Sk~-69~202, all
of it should be made available.  We report on an ultralow--dispersion
objective--prism spectrum of Sk~-69~202 and the surrounding region
that was obtained in January 1974.  Formally, this spectrum extends the
wavelength coverage of Sk~-69~202 to 6600~\AA, although only at
ultralow--dispersion.

\section{ULTRALOW--DISPERSION SPECTRUM OF Sk~-69~202}

Morgan et~al. (1954) showed that faint OB stars and very red stars
could be easily detected on 103a--F plates using objective prisms that
give ultralow dispersion of 30,000~\AA~mm$^{-1}$. Subsequent work by
Schulte (1956a,b) identified several reddened OB stars of the
Cygnus~OB2 (VI~Cygni) association from 10,000~\AA~ mm$^{-1}$
objective--prism spectra. This technique requires construction of
prisms of very small angle (Bidelman 1972). The ultralow--dispersion
technique was developed at the Vainu Bappu Observatory (VBO), Kavalur,
India and was successfully applied to detect red and blue stars (Bappu
\& Parthasarathy 1977; Parthasarathy 1978). Bappu, Parthasarathy, \&
Scaria (1977, 1978, 1985) used this method to detect very red stars in
selected regions of the LMC, and Parthasarathy \& Jain (1995) applied
it to detect cool and reddened stars in Cyg~OB2.

The instrument (Bappu \& Parthasarathy 1977) used was an F/2 slitless
spectrograph with a three--degree quartz prism and a Schmidt camera at
the Cassegrain focus of the 1--m Ritchey--Chretien reflector at
VBO. The field of view was 40 arcminutes in diameter.  Spectra
covering the wavelength range 3500 to 6600~\AA\ were obtained on Kodak
103a-E emulsion.  The dispersion was 10,000~\AA~mm$^{-1}$. The spectra
were unwidened and were about 250 microns in length along the
dispersion (Bappu \& Parthasarathy 1977; Parthasarathy \& Jain
1995). With this setup ultralow--dispersion spectra of all stars
within a 40 minutes of arc field could be recorded in one exposure.

Ultralow--dispersion spectra of B, A, F, G, K, and M stars are shown
in Figure~1. The main criterion for classification is the shape of the
image. From Figure~1 it is clear that one can easily distinguish
between OB stars and very red stars. The discontinuity in the spectrum
is due to the dip in the sensitivity of the 103a--E emulsion in the
green, which serves as a wavelength reference to distinguish the blue
part of the spectrum from the red. For example notice the difference
in the spectra of the B star and the K and M stars.  Observations of
several selected regions of the LMC obtained between 1974 and 1976 and
the detection of red stars in the LMC can be found in Bappu
et~al. (1977, 1978, 1985).  The region of the LMC observed in 1974 in
which Sk~-69~202 is present is shown in Figure~2. This region also
includes the nearby Sk~-69~203, a B0.7~I star with $V = 12.29, B-V =
+0.01, U-B = -0.85$ (Isserstedt 1975; Rousseau et~al. 1978; Walborn
et~al. 1989), quite similar to Sk~-69~202 and possibly associated with
it.

The ultralow--dispersion spectrum of Sk~-69~202 shows that it was a B
star in 1974.  Comparison of densities in different sections of the
images of Sk~-69~202 and Sk~-69~203 indicates that the magnitude and
colors of Sk~-69~202 were similar to that of Sk~-69~203, in agreement
with the UBV photometry and spectral classification.

\section{PRESUPERNOVA EVOLUTION OF Sk~-69~202}

All data discussed above indicate that from 1971 to 1983 Sk~-69~202
showed the characteristics of a normal B3~I star, with no evidence for
variability, no Balmer--line emission, and no noticeable reddening or
near--IR excess. (Walborn et~al. 1989). From the spectra presented by
Walborn et~al. we estimate that the rotational velocity may not be
more than 100 km s$^{-1}$.  Sk~-69~202 was a normal B3~I supergiant
shortly before exploding.

SN~1987A has three circumstellar rings that are part of a complex
system of gas that appears to have been ejected by the progenitor some
20,000 years before Sk~-69~202 exploded (Burrows et al. 1995; Sugerman
et~al. 2005 and many references therein).  The inner, equatorial ring
is expanding at 10~km~s$^{-1}$; its true shape is a nearly circular
ring of radius~0.67 ly, inclined at 43 degrees to our line of sight.
Analysis of the emission--line spectrum of the inner ring (Fransson
et~al. 2005) indicates that it is nitrogen-rich, with N/(C+O) about 10
times greater than the solar ratio (or perhaps somewhat smaller than
this; Zhekov et~al. 2006). The enhanced N abundance is thought to be
due to conversion of C and O to N by the CNO cycle in Sk~-69~202
before the ring material was expelled.  The outer loops have physical
radii 1.5~ly and lie on planes approximately parallel to the plane of
the inner ring but displaced by about 1.3~ly on either side. The outer
loops have radial velocities consistent with the interpretation that
they were expelled at or near the time of the inner ring material.

The presupernova evolution of Sk~-69~202 and the origin of the triple
ring system remain uncertain.  According to the evolutionary models of
massive stars the progenitors of most SNe II are expected to be RSGs.
The prevailing opinion is that SN~1987A did become an RSG, during
which time it may have expelled the material of the rings in the form
of a slow wind, and then returned to the blue. Within the current
uncertainties of stellar evolution it is possible to make assumptions
that cause models of single stars to return to the blue, but it is not
clear that these assumptions are natural (Woosley et~al. 2002).  It is
supposed that the fast wind of the second BSG phase then interacted
with the slow wind of the RSG phase to produce the present ring system
(e.g., Lundqvist \& Fransson 1996; Sugarman et~al. 2005).  However, to
end up with a ring system rather than a more spherical shell, the RSG
wind must have been for some reason equatorially enhanced, and if
Sk~-69~202 had been an RSG one might expect some evidence for
circumstellar gas and dust, emission lines, and variability in the
presupernova data obtained between 1971 and 1983.

Opposed to the single--star scenario above is that Sk~-69~202 may have
been the product of the merger of an RSG and a less massive companion
star (Hillebrandt \& Meyer 1989; Podsiadlowski \& Ivanova 2003).  Such
a merger can result in a BSG and perhaps a complex circumstellar
system.  However, the scenario is complicated.  Morris \&
Podsiadlowski (2006) find that the mass ejection would be primarily at
mid--latitudes rather than in the equatorial plane.  They suggest that
the mid--latitude ejection provided the material of the outer rings,
and that the equatorial ejection of the inner ring material occurred
{\sl after} the merger in a rotation--enforced outflow during the
contraction toward the blue. Another possible difficulty with the
merger model is that it requires fine tuning to get the merger to
happen just 20,000 years before the supernova, which would make
SN~1987A an uncommon event (Woosley et~al. 2002).  Also, if Sk~-69~202
was a merger product, rapid rotation of a few hundred km s$^{-1}$ is
expected. As mentioned above, the spectra of Sk~-69~202 showed no
evidence for rapid rotation.

Is it possible that Sk~-69~202 never was an RSG?  Pastorello
et~al. (2005) presented spectroscopic and photometric observations of
the peculiar Type~II SN~1998A. They showed that in many respects the
light curves and spectra resembled those of SN~1987A, indicating that
the progenitor of SN~1998A also was a BSG.  In spite of the strong
observational selection against the discovery of subluminous events,
Pastorello et~al. cited the existence of several other SNe~II with
(fragmentary) light curves possibly similar in shape to SN~1987A and
SN~1998A, which suggests that explosions of BSGs are not too uncommon.
This would be inconsistent with explanations that require fine tuning.
Instead, there may be some regular channel for producing
SN~1987A--like events.  From stellar evolution calculations H\"oflich
et~al. (2001) found that models having somewhat lower metallicity or
higher main---sequence mass than SN~1987A explode as BSGs without
becoming RSGs, but the quantitative results depend on various
assumptions.  A problem for this scenario for SN~1987A, though, is
that if Sk~-69~202 never was an RSG, then there may be no good
explanation for the low--velocity circumstellar material, although
McCray \& Lin (1994) suggested that the inner ring is the inner rim of
a disk of gas left over from the formation of Sk~-69~202.  It is
interesting that Sher~25, an apparently normal B1.5~Iab supergiant in
or near the Galactic open cluster NGC~3603, has a system of rings
reminiscent of SN~1987A (Brandner et~al. 1997).  Like other B~type
supergiants, Sher~25 has photospheric CNO abundances that indicate
strong processing via the CNO cycle (Crowther et~al. 2007).

The history of Sk~-69~202 continues to be a mystery.  Future
observations of the interaction between the high--velocity ejecta of
SN~1987A and the circumstellar matter, especially in X--rays (Zhekov
et~al. 2006), may provide vital clues to the past.

We are grateful to Arlin Crotts and Dick McCray for instructive
comments.  This work has been supported by NSF grant AST-0506028 and
NASA grant NNG04GD368.

%
%
%
%
%
%
%
%
%

\clearpage

\begin{figure}
\begin{center}
\includegraphics[height=.8\textwidth]{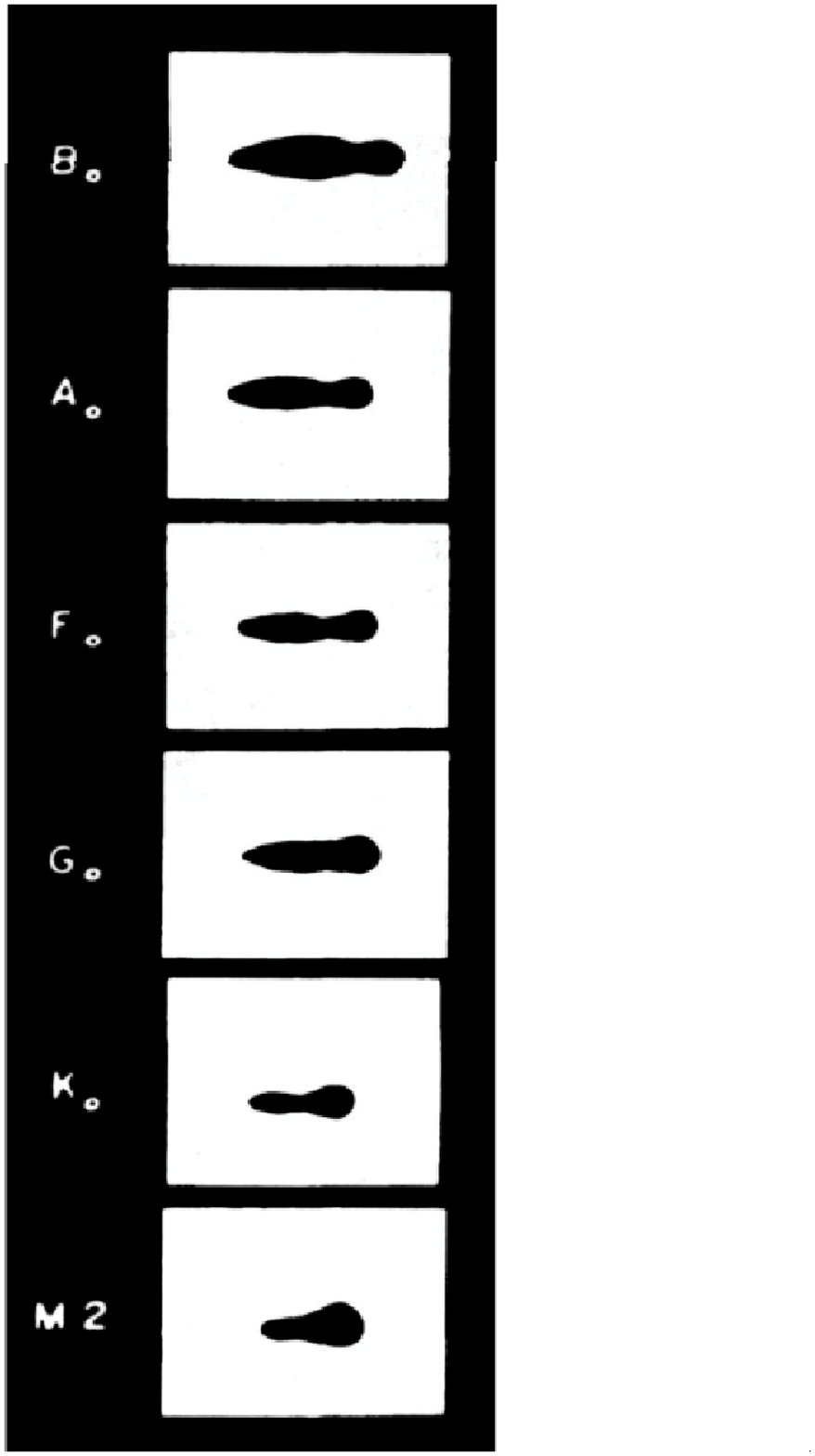}
\end{center}
\caption{The ultralow--dispersion (10,000~\AA~mm$^{-1}$)
spectra of stars of various spectral types.  Wavelength increases to
the right.  The discontinuity in the spectrum is due to the dip in the
sensitivity of the 103a--E emulsion in the green.
}
\end{figure}

\begin{figure}
\begin{center}
\includegraphics[width=.8\textwidth]{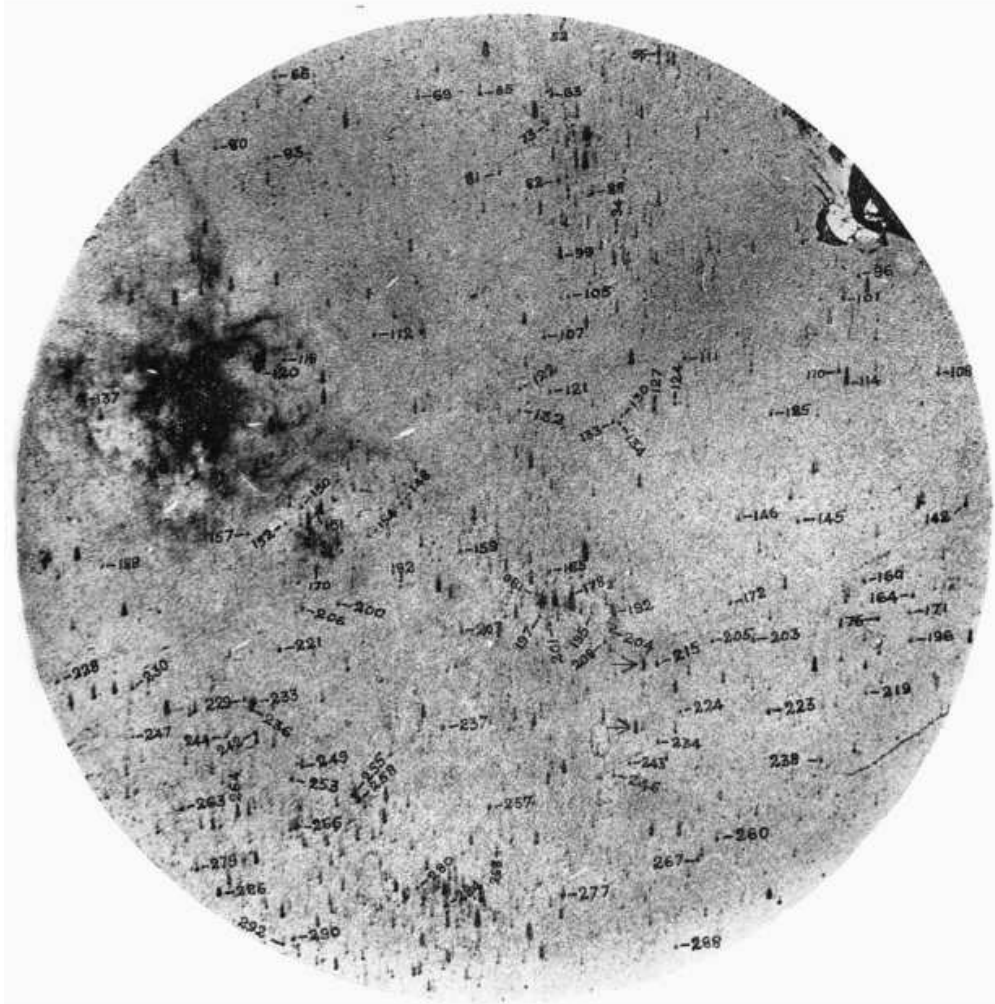}
\end{center}
\caption{
The region of the LMC that contains Sk -69 202
(lower arrow), the progenitor of SN~1987A, and Sk -69 203 (upper
arrow).  Numbered stars are red stars identified by Bappu
et~al. (1977).  Spectra were recorded on 103a-E emulsion and cover the
wavelength range 3500 to 6500~\AA.  Wavelength increases downward.
The 1950 coordinates of the field center are R.A. 05 36.5 and Dec. -69
10.  North is at the top and east is toward the left.
}
\end{figure}

\end{document}